\documentclass[10pt,aps,prd,floatfix,notitlepage,nofootinbib]{revtex4-1}
\usepackage[utf8]{inputenc}
\usepackage{multirow}
\usepackage{latexsym}
\usepackage{epsfig,graphics,subfig,float}
\usepackage{epstopdf}

\usepackage{tabularx,booktabs,caption}

\usepackage{amssymb}
\usepackage{verbatim}
\usepackage{multirow}
\usepackage{color}
\usepackage{tikz}
\usepackage{amsmath,amssymb,amsfonts}

\usepackage{tikz}
\usepackage{bm}
\usetikzlibrary{matrix}

\newcommand{\udt}[3]{#1^{#2}_{\phantom{#2}#3}}

\newcommand{\dut}[3]{#1_{#2}^{\phantom{#2}#3}}

\begin{document}

\title{Generalized Tachyonic Teleparallel cosmology}
\author{Sebastian Bahamonde}
\email{sbahamonde@ut.ee,sebastian.beltran.14@ucl.ac.uk}
\affiliation{Laboratory of Theoretical Physics, Institute of Physics, University of Tartu, W. Ostwaldi 1, 50411 Tartu, Estonia}
\affiliation{Department of Mathematics, University College London,
	Gower Street, London, WC1E 6BT, United Kingdom}
    \affiliation{   School of Mathematics and Physics, University of Lincoln.
   	Brayford Pool, Lincoln, LN6 7TS, United Kingdom}
    \affiliation{University of Cambridge, Cavendish Laboratory, JJ Thomson Avenue, Cambridge CB3 0HE, United Kingdom}

\author{Mihai Marciu}
\email{mihai.marciu@drd.unibuc.ro}
\affiliation{Faculty of Physics, University of Bucharest, 405 Atomi\c{s}tilor, POB MG-11, RO-077125 Bucure\c{s}ti-M\u{a}gurele, Rom\^{a}nia}

\author{Jackson Levi Said}
\email{jackson.said@um.edu.mt}
\affiliation{Institute of Space Sciences and Astronomy, University of Malta, Msida, MSD 2080, Malta}
\affiliation{Department of Physics, University of Malta, Msida, MSD 2080, Malta}

\begin{abstract}
In this paper we propose a new dark energy model in the teleparallel alternative of general relativity, by considering a generalized non--minimal coupling of a tachyonic scalar field with the teleparallel boundary term. Within the framework of teleparallel gravity, the boundary coupling term is associated with the divergence of the torsion vector. Considering the linear stability technique for various potentials and couplings, we have analyzed the dynamical properties of the present tachyonic dark energy model in the phase space, uncovering the corresponding essential dynamical features. Our study of the phase space structure revealed that for a specific class of potential energy, this model exhibits various critical points which are related to different cosmological behaviors, such as accelerated expansion and scaling solutions, determining the existence conditions and the corresponding physical features.

\end{abstract}

\date{\today}

\maketitle



\section{Introduction}\label{sec:1}
Einstein's theory of General Relativity (GR) has experienced unprecedented success in its power to explain astrophysical phenomena ranging from Solar System tests to strong field gravitational wave physics \cite{Misner:1974qy}. However, this theory of gravity has required important modifications due to observational realities which have arisen over the past few decades. In terms of the energy budget of the Universe, the first modification comes from observations of galaxies and their dynamical structure, which is only possible with the addition of approximately purely gravitational interacting particles, namely \textit{dark matter}, which may potentially be beyond the standard model of particle physics \cite{Rubin:1970zza,Navarro:1995iw}. The second and larger contribution to the modification of GR comes from the relatively recent observation of the accelerating expansion of the Universe \cite{Riess:1998cb,Perlmutter:1998np} which is an observational fact, called \textit{dark energy}. This can be accounted for in GR through the introduction of the cosmological constant however this poses its own problems \cite{Weinberg:1988cp}. The $\Lambda$CDM is the most successful model which can explain the current accelerated expansion and the evolution of the observable Universe at the level of background dynamics, involving the superposition between the dark matter fluid and the cosmological constant. On the other hand, the early period of the Universe also features several facets that need remedy. Most prominently, for $\Lambda$CDM to correctly produce our current picture of the Universe, a period of cosmological inflation must of taken place \cite{Guth:1980zm,Linde:1981mu} which would allow for a natural solution to the horizon problem. However, this may also necessitate further particles beyond the standard model \cite{Lyth:1998xn}. Cosmologically, the time that should be best described by the $\Lambda$CDM model is the present or late-time period of the Universe. However, recent releases by the Planck collaboration have revealed a growing tension in the local and global measurements of $H_0$ and $f\sigma_8$ \cite{Aghanim:2018eyx}. \medskip

Now, it may be the case that the fundamental and observational problems surrounding $\Lambda$CDM may be resolved in the coming years, or it may be the case that $\Lambda$CDM needs to be changed in some way. Over the previous decades there has been concerted efforts in extending GR to account for certain elements of these problems \cite{Capozziello:2011et}. However, it may also be the case that a new paradigm is needed to confront the growing requirements of constructing a viable theory of gravity. One such treatment is the teleparallel gravity approach where the Levi-Civita connection is replaced with the Weitzenb\"{o}ck connection \cite{AP}. The connection plays a crucial role in gravitational physics in that the expression of curvature, torsion, or nonmetricity is not a property of the manifold itself but of the connection which relates the elemental tangent spaces of the manifold \cite{Heisenberg:2018vsk}. In this way, one can choose to consider gravitation in terms of the curvatureless Weitzenb\"{o}ck connection which also observes the metricity property. \medskip

In teleparallel gravity, the gravitation is characterized by the torsion tensor, $\udt{T}{\rho}{\mu\nu}$, instead of the Riemann tensor in GR and its extensions. As in the GR framework, a Lagrangian can be constructed to represent the gravitational field. Of particular interest, in teleparallel gravity, is that a Lagrangian can be constructed such that it is equivalent to the Einstein-Hilbert Lagrangian up to a total divergence or boundary term, $B$, that is
\begin{equation}\label{ricc_tor_sca_def}
    R=-T+B,
\end{equation}
where $T$ is called the torsion scalar and contains only second order terms, while the boundary term, $B$, encapsulates the higher order contributions to the Ricci scalar, $R$. This is the so-called \textit{teleparallel equivalent of general relativity} (TEGR) which is equivalent to GR at the level of the field equations \cite{Cai:2015emx,Krssak:2018ywd}. The natural consequence of this realization is that every test of GR also becomes a test of TEGR with the difference that in TEGR gravity is again observed to act as a (Lorentz) force, and that the barrier with the quantum regime seems to have less tension \cite{AP}. Moreover, due to the second order nature of the torsion tensor, even in the case where the theory is extended to an $f(T)$ Lagrangian, the resulting field equations remain second order which has important consequences for the gravitational wave polarization modes of the theory. In fact, $f(T)$ gravity continues to exhibit the equivalent polarization modes as in the GR and TEGR settings \cite{Farrugia:2018gyz}. \medskip

Given the decomposition of the Ricci scalar into the second and fourth order terms expressed in Eq.~(\ref{ricc_tor_sca_def}), we consider the analysis of a tachyonic dark energy model nonminimally coupled to the aforementioned separate contributions through different functionals. In particular, we choose to study a tachyonic scalar field which has been shown to produce an inflationary epoch and late-time accelerating solutions that do not violate the strong energy condition \cite{Quiros:2009mz,Aguirregabiria:2004xd}. These models are partially inspired by string theory \cite{Bagla:2002yn,Fang:2010zze,Ortin:2015hya} and $k$-essence theory \cite{Almeida:2016ixq}. In the teleparallel setting, scalar fields have been investigated to a moderate degree, with various extensions having been investigated in the cosmological context \cite{Geng:2011aj,Xu:2012jf,Hohmann:2018rwf,PhysRevD.97.084008,Otalora:2013tba,Otalora:2014aoa}. Furthermore, the effects of the boundary couplings in scalar tensor theories have been discussed in various papers \cite{Hohmann:2018ijr,Bahamonde:2018miw,Abedi:2018lkr,Hohmann:2018rwf, Bahamonde:2017wwk, Marciu:2017sji,Gecim:2017hmn, Bahamonde:2017ifa, Abedi:2017jqx, Fazlpour:2018jzz,Bahamonde:2016grb, Bahamonde:2016cul,MohseniSadjadi:2016ukp, Bahamonde:2015zma}. A recent review on various studies related to dynamical analysis in different cosmological constructions can be found in \cite{BAHAMONDE20181}. In Ref.~\cite{Otalora:2013dsa}, a tachyonic field is investigated for the modified teleparallel setting, late-time accelerating attractor solutions are found with a field equation of state that realistically tends to the current dark energy value. This approach was extended to more general models in Ref.~\cite{Fazlpour:2014qaa} where the attractor solution context is further clarified.    \medskip

The work is divided as follows: in Sec.~\ref{sec:2} the tachyonic approach to extended teleparallel theories of gravity is introduced with a focus on the cosmological consequences of the treatment. In Sec.~\ref{sec:3}, the dynamical analysis of the system is undertaken for specific choices of tachyonic field. Finally, in Sec.~\ref{sec:4} the conclusions are summarized and discussed. Unless stated otherwise geometric units are used throughout the paper. In addition, $e^a_\mu$ and $E_a^\mu$ represent the tetrads and the inverse of the tetrads respectively and the $(+---)$ metric signature is used.

\section{Generalized Tachyonic Teleparallel theories of gravity}\label{sec:2}
\noindent In this paper, we present a new teleparallel tachyonic model based on the following action
\begin{equation}
S = \int
\left[\frac{T}{2\kappa^2}+\frac{1}{2}f(\phi) T+\frac{1}{2}g(\phi)B-V(\phi)\sqrt{1-\frac{2X}{V(\phi)}}+L_{\rm m}\right] e\, d^4x \,,\label{1}
\end{equation}
where $\kappa^2=8\pi G$, $L_{\rm m}$ is a matter Lagrangian, $T$ is the scalar torsion, $B=(2/e)\partial_{\mu}(eT^{\mu})$ is the boundary term, $f(\phi)$ and $g(\phi)$ are scalar field dependent coupling functions, $V(\phi)$ is the potential
and
\begin{equation}
    X=\frac{1}{2}(\partial_\mu \phi)( \partial^\mu \phi)
\end{equation}
is the kinetic term. The torsion tensor is identified as
\begin{equation}
    \udt{T}{a}{\mu\nu}=\partial_{\mu}e^a_{\nu}-\partial_{\nu}e^a_{\mu}+\udt{\omega}{a}{b\mu}e^b_{\nu}-\udt{\omega}{a}{b\nu}e^b_{\mu}\,,
\end{equation}
where $e^a_{\nu}$ form a tetrad field of the gravitational system and represent coordinate transformations between the general manifold and the tangent space at any point, while $\udt{\omega}{a}{b\mu}$ form the spin connection components which are purely inertial and sustain the local Lorentz invariance of the theory \cite{Krssak:2015oua,Cai:2015emx}. The torsion scalar is then defined through the contraction
\begin{equation}
    T=\udt{T}{a}{\mu\nu}\dut{S}{a}{\mu\nu},
\end{equation}
where the superpotential is defined as 
\begin{equation}
    \dut{S}{a}{\mu\nu}=\frac{1}{2}\left(\dut{T}{a}{\mu\nu}+\udt{T}{\nu\mu}{a}-\udt{T}{\mu\nu}{a}\right)-E^a_{\nu}\udt{T}{\alpha\mu}{\alpha}+E^a_{\mu}\udt{T}{\alpha\nu}{\alpha}\,.
\end{equation}
This work considers an analogous generalization of other Tachyonic models studied in the literature \cite{Otalora:2013dsa}. By taking, $f(\phi)=-g(\phi)$, one recovers a Tachyonic theory with a non-minimally coupling between the scalar field and the Ricci scalar $R$ due to the relation in Eq.~(\ref{ricc_tor_sca_def}). By taking $g(\phi)=0$, one recovers a teleparallel Tachyonic theory where the torsion scalar $T$ is non-minimally coupled with the scalar field. In \cite{Banijamali:2012nx}, the authors found that this coupling allows the crossing of the phantom divide line. The new coupling between the scalar field and the boundary term is motivated from the scalar tensor theory studied in \cite{Bahamonde:2015hza,Zubair:2016uhx}, where the authors found that, without fine-tunning, the system evolves to a late-time accelerating attractor solution.

\noindent By taking variation with respect to the tetrad, we find the corresponding gravitational field equations given by
\begin{eqnarray}
2\Big(\frac{1}{\kappa^2}+f(\phi)\Big)\left[ e^{-1}\partial_\mu (e S_{a}{}^{\mu\nu})-E_{a}^{\lambda}T^{\rho}{}_{\mu\lambda}S_{\rho}{}^{\nu\mu}-\frac{1}{4}E^{\nu}_{a}T\right]-E_a^\nu V(\phi)\sqrt{1-\frac{2X}{V(\phi)}}
\nonumber\\ -\frac{1}{\sqrt{1-\frac{2X}{V(\phi)}}}E_a^\nu \partial_\mu \phi \partial^\mu \phi+ 2(\partial_{\mu}f(\phi)+\partial_{\mu}g(\phi)) E^\rho_a S_{\rho}{}^{\mu\nu}+E^{\nu}_{a}\Box g(\phi)-E^\mu_a \nabla^{\nu}\nabla_{\mu}g(\phi)=  \mathcal{T}^\nu_a\,. \label{2}
\end{eqnarray}
Here, $\nabla_\mu$ is the covariant derivative with respect to the Levi-Civita connection and $\Box=\nabla^\mu\nabla_\mu$. By taking variations with respect to the scalar field, one finds
\begin{align}
  \frac{V'(\phi)}{2V(\phi)\sqrt{1-\frac{2X}{V(\phi)}}}(\partial_\mu \phi)( \partial^\mu \phi)   +\partial_\mu \Big(\frac{\partial^\mu \phi}{\sqrt{1-\frac{2X}{V(\phi)}}}\Big)+V'(\phi)\sqrt{1-\frac{2X}{V(\phi)}}&=\frac{1}{2}\Big(f'(\phi)T+g'(\phi)B\Big)\,.\label{KG}
\end{align}

For the flat FLRW metric in Cartesian coordinates, the metric is given by
\begin{equation}
    ds^2=dt^2-a(t)^2(dx^2+dy^2+dz^2)\,,
\end{equation}
where $a(t)$ is the cosmological scale factor. This can equivalently be described by the tetrad field
\begin{equation}
    e_\mu^a=\textrm{diag}(1,a(t),a(t),a(t))\,,
\end{equation}
which naturally induce vanishing spin connection components \cite{Krssak:2015oua,Cai:2015emx}. The ensuing Friedmann equations then turn out to be represented by
\begin{eqnarray}
3H^2&=&\rho_{\rm m}-3H^2 f(\phi)+3 H \dot{\phi}g'(\phi )+\frac{V(\phi )}{\sqrt{1-\frac{\dot{\phi}^2}{V(\phi)}}}\,,\label{eq1}\\
3 H^2+2 \dot{H}&=&-p_{\rm m}-3 H^2f(\phi)-2 f(\phi) \dot{H}-2  H \dot{\phi}f'(\phi)+\dot{ \phi}^2 g''(\phi)+  \ddot{ \phi} g'(\phi)+V(\phi) \sqrt{1-\frac{\dot{\phi}^2}{V(\phi)}}\,,\label{eq2}
\end{eqnarray}
where $\kappa=1$ was assumed, dots represent differentiation with respect to cosmic time and primes with respect to the scalar field. It should be noted that a standard perfect fluid with energy density $\rho_{\rm m}$ and pressure $p_{\rm m}$ is being assumed. It is straightforward to show that the standard conservation of the energy-momentum tensor gives 
\begin{align}
    \dot{\rho}_{\rm m}+3H(\rho_{\rm m}+p_{\rm m})=0\,,
\end{align}
which is the standard conservation equation for matter.

\noindent The scalar field relation in Eq.~(\ref{KG}) takes the following form
\begin{eqnarray}
\ddot{\phi}+3 H\dot{\phi} \left(1-\frac{\dot{\phi}^2}{V(\phi )}\right)+\left(1-\frac{3 \dot{\phi}^2}{2 V(\phi)}\right) V'(\phi)&=&-\left(1-\frac{\dot{\phi}^2}{V(\phi )}\right)^{3/2}\left[3 H^2 f'(\phi)+3 \left(3 H^2+\dot{H}\right) g'(\phi) \right]\,.
\end{eqnarray}
It can be shown that this equation can be also found directly from the modified FLRW equations \eqref{eq1}--\eqref{eq2}, therefore, it is not an extra constraint equation.

\noindent The Friedmann equations can also equivalently be rewritten as their GR analogue with an additional effective fluid component so that
\begin{eqnarray}
3H^2=\rho_{\rm eff}\,,\quad 3 H^2+2 \dot{H}=-p_{\rm eff}\,,
\end{eqnarray}
where $\rho_{\rm eff}=\rho_{\rm m}+\rho_{\phi}$ and $p_{\rm eff}=p_{\rm m}+p_{\phi}$ with the new fluid quantities defined as follows
\begin{align}
    \rho_{\phi}&=-3H^2 f(\phi)+3 H \dot{\phi}g'(\phi )+\frac{V(\phi )}{\sqrt{1-\frac{\dot{\phi}^2}{V(\phi)}}}\,,\\
    p_{\phi}&=3 H^2f(\phi)+2 f(\phi) \dot{H}+2  H \dot{\phi}f'(\phi)-\dot{ \phi}^2 g''(\phi)-  \ddot{ \phi} g'(\phi)-V(\phi) \sqrt{1-\frac{\dot{\phi}^2}{V(\phi)}}\,.
\end{align}
It is then convenient to introduce the effective equation of state parameter
\begin{eqnarray}
w_{\rm eff}=\frac{p_{\rm eff}}{\rho_{\rm eff}}=\frac{p_{\rm m}+p_{\phi}}{\rho_{\rm m}+\rho_{\phi}}\,,
\end{eqnarray}
which means that the density parameters of the contributing components take the form
\begin{align}
    \Omega_{\rm m}=\frac{\rho_{\rm m}}{3H^2}\,,\quad \Omega_{\phi}=\frac{\rho_{\phi}}{3H^2}\,,\label{energydensity}
\end{align}
in such a way that $\Omega_{\rm m}+\Omega_{\phi}=1$ holds.

\section{Dynamical system of the model}\label{sec:3}
In this section, we will study the dynamical system of our model which only has non-minimally couplings between the scalar field and the boundary terms, therefore we will assume that $f(\phi)=0$, and $g(\phi)\neq0$. The dynamical system for the case $f(\phi)\neq 0$ and $g(\phi)=0$ was studied previously in~\cite{Otalora:2013dsa,Fazlpour:2014qaa}. Let us first introduce the following dimensionless variables
\begin{align}
    x=\frac{\dot{\phi}}{\sqrt{V(\phi)}}\,,\quad y=\frac{\sqrt{V(\phi)}}{\sqrt{3}H}\,,\quad u=\frac{1}{2}g'(\phi)\,,\quad \lambda=-\frac{V'(\phi)}{V(\phi)}\,,
\end{align}
so that $y>0$. By using these variables in the first FLRW equation in Eq.~(\ref{eq1}) and the definition of the energy density parameter given by Eq.~(\ref{energydensity}), one gets the constraint
\begin{align}
     0 \leq \Omega_{\rm m}= 1-2 \sqrt{3}   u x y-\frac{y^2}{\sqrt{1-x^2}}\leq 1\,,
\end{align}
which gives the phase space of the dynamical system. One needs to choose $g(\phi)$ to write down the full dynamical system of the model. Therefore, in the next sections, we will study two different kind of couplings: power-law and exponential types. We will further assume standard a barotropic fluid given by $p_{\rm m}=w_{\rm m} \rho_{\rm m}$.

\subsection{Power-law coupling and exponential potential}
Assuming a power-law coupling between the boundary term and the scalar field  means setting
\begin{eqnarray}
g(\phi)=\chi \phi^p\,, 
\end{eqnarray}
where $\chi$ and $p$ are both constants. It is possible to write down the dynamical equations of the system as a 4-dimensional one with a generic potential. This can be done if one introduces the variable 
\begin{align}
    \Gamma=\frac{V(\phi) V''(\phi)}{V'(\phi ) V'(\phi)}\,.
    \label{ecuatie_gamma}
\end{align}
By introducing the variable $N=\log(a)$, the dynamical system for this kind of couplings can be written as
\begin{equation}
\frac{dx}{dN}=\frac{\ddot{\phi}}{\sqrt{3} H^2 y}+\frac{1}{2} \sqrt{3} \lambda  x^2 y, \nonumber
\end{equation}
\begin{equation}
    \frac{dy}{dN}=-\frac{y \dot{H}}{H^2}-\frac{1}{2} \sqrt{3} \lambda  x y^2, \nonumber
\end{equation}
\begin{equation}
\frac{du}{dN}=\sqrt{3} \cdot 2^{\frac{1}{1-p}} (p-1) p \chi  x y \left(\frac{u}{p \chi }\right)^{\frac{p-2}{p-1}}, \nonumber
\end{equation}
\begin{equation}
\frac{d\lambda}{dN}=-\sqrt{3} (\Gamma-1) \lambda^2 x y\,,
\nonumber
\end{equation}
where
\begin{equation}
    \ddot{\phi}=3 H^2 \lambda  \left(1-\frac{3 x^2}{2}\right) y^2-3 \sqrt{3} H^2 x \left(1-x^2\right) y-6 u \left(1-x^2\right)^{3/2} \left(3 H^2+\dot{H}\right),\nonumber
\end{equation}
and
\begin{multline}
    \dot{H}=\frac{3 H^2}{12 u^2 \left(1-x^2\right)^{3/2}+2}\cdot \Bigg[ w_m \left(2 \sqrt{3} u x y+\frac{y^2}{\sqrt{1-x^2}}-1\right)+2^{\frac{p-2}{p-1}} (p-1) p \chi  x^2 y^2 \left(\frac{u}{p \chi }\right)^{\frac{p-2}{p-1}}
    \\
    -12 u^2 \left(1-x^2\right)^{3/2}-\lambda  u \left(3 x^2-2\right) y^2+2 \sqrt{3} u x \left(x^2-1\right) y+\sqrt{1-x^2} y^2-1 \Bigg].\label{eq:11}
\end{multline}
This dynamical system assumes that $p\neq 1$. The case $p=1$ is a very special one which gives a 3-dimensional dynamical system of equations, but generically, does not give any interesting cosmological behaviour. One more realistic model is the one where $p=2$ which gives a coupling like $(1/2)\chi \phi^2 B $ which was studied in \cite{Bahamonde:2015hza}. In that case however, non-tachyonic scalar fields were considered. In what follows, we shall consider the case where we have a specific coupling, $g(\phi)=\chi \phi^p$, with $p=2$ and an exponential potential $V(\phi)=V_0 e^{- \lambda \phi}$, where $\lambda$ is a positive constant. In this case, the dynamical evolution of the model can be described by a 3D autonomous system of differential equations
\begin{multline}
\label{equu1}
    \frac{dx}{dN}=\frac{1}{y \left(6 u^2 \left(x^2-1\right)^2+\sqrt{1-x^2}\right)} \Big(  -18 u^2 x^5 y w_{\rm m}+36 u^2 x^3 y w_{\rm m}-18 u^2 x y w_{\rm m}+3 \sqrt{3} u x^4 w_{\rm m}+3 u \sqrt{3-3 x^2} x^2 y^2 w_{\rm m}
    \\-3 u \sqrt{3-3 x^2} y^2 w_{\rm m}-6 \sqrt{3} u x^2 w_{\rm m}+3 \sqrt{3} u w_{\rm m}+3 \sqrt{3} \lambda  u^2 x^6 y^2-6 \sqrt{3} \lambda  u^2 x^4 y^2+3 \sqrt{3} \lambda  u^2 x^2 y^2
    \\-6 \sqrt{3} u \chi  x^6 y^2+12 \sqrt{3} u \chi  x^4 y^2-3 \sqrt{3} u x^4-6 \sqrt{3} u \chi  x^2 y^2
    +6 \sqrt{3} u \sqrt{1-x^2} x^2 y^2-3 \sqrt{3} u \sqrt{1-x^2} y^2
    \\+6 \sqrt{3} u x^2-3 \sqrt{3} u \sqrt{1-x^2} x^4 y^2-3 \sqrt{3} u-\sqrt{3} \lambda  \sqrt{1-x^2} x^2 y^2+\sqrt{3} \lambda  \sqrt{1-x^2} y^2-3 \sqrt{1-x^2} x y+3 \sqrt{1-x^2} x^3 y\Big) \,,
\end{multline}
\begin{multline}
    \frac{dy}{dN}=\frac{1}{2 \left(6 u^2 \left(x^2-1\right)^2+\sqrt{1-x^2}\right)} \Big( -6 u \sqrt{3-3 x^2} x y^2 w_{\rm m}+3 \sqrt{1-x^2} y w_{\rm m}-3 y^3 w_{\rm m}-6 \sqrt{3} \lambda  u^2 x^5 y^2+36 u^2 x^4 y
    \\+12 \sqrt{3} \lambda  u^2 x^3 y^2-72 u^2 x^2 y-6 \sqrt{3} \lambda  u^2 x y^2+36 u^2 y+9 \lambda  u \sqrt{1-x^2} x^2 y^3-6 \lambda  u \sqrt{1-x^2} y^3
    \\+6 \sqrt{3} u \sqrt{1-x^2} x y^2-6 \sqrt{3} u \sqrt{1-x^2} x^3 y^2-6 \chi  \sqrt{1-x^2} x^2 y^3+3 x^2 y^3-\lambda  \sqrt{3-3 x^2} x y^2+3 \sqrt{1-x^2} y-3 y^3 
    \Big)\,,
\end{multline}
\begin{equation}
\label{equu2}
    \frac{du}{dN}=\sqrt{3} \chi  x y.
\end{equation}
\par
\begin{figure}[H]
    \begin{minipage}{0.45\textwidth}
        \centering
        \includegraphics[width=0.9\textwidth]{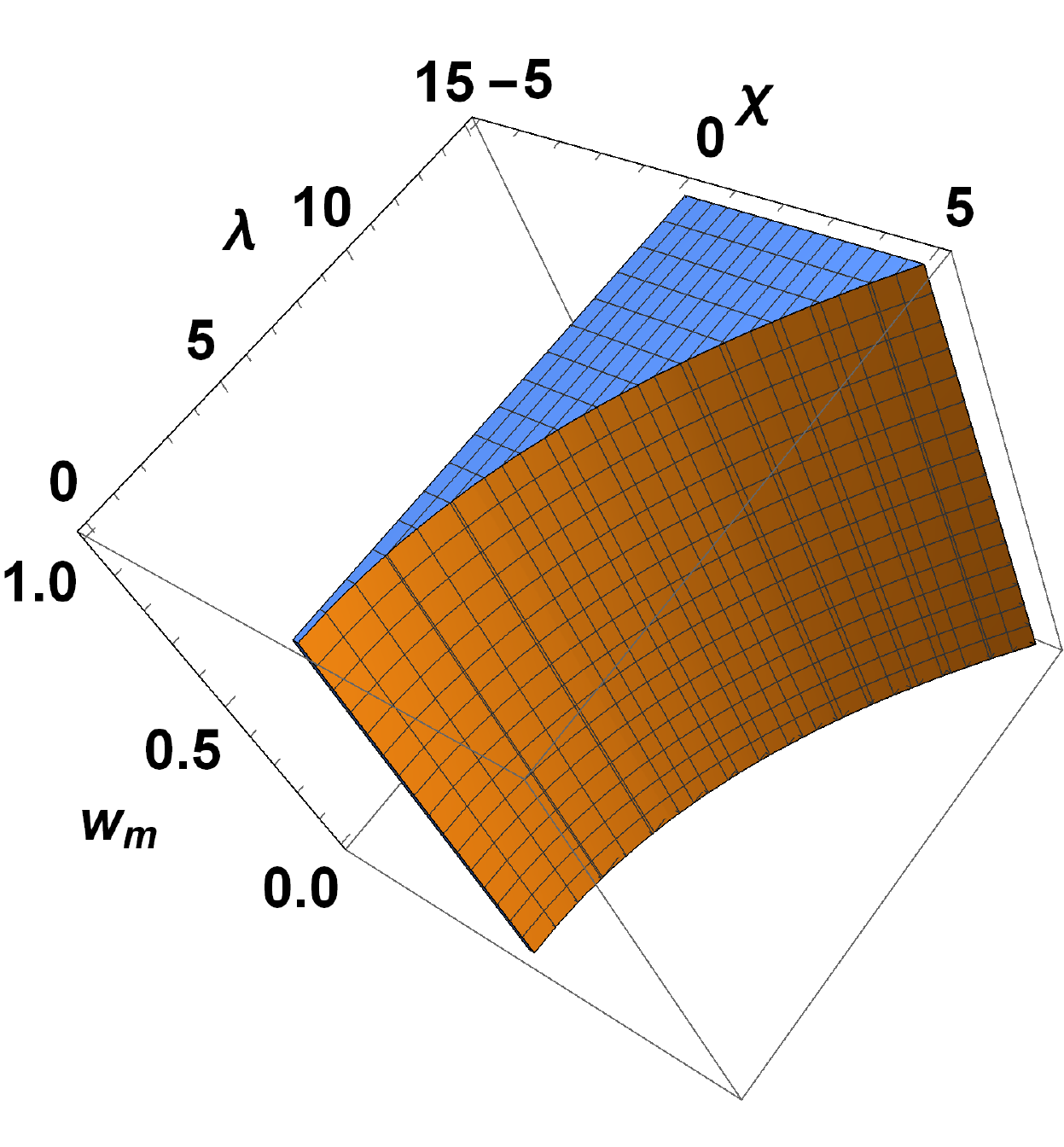} 
    \end{minipage}\hfill
    \begin{minipage}{0.45\textwidth}
        \centering
        \includegraphics[width=0.9\textwidth]{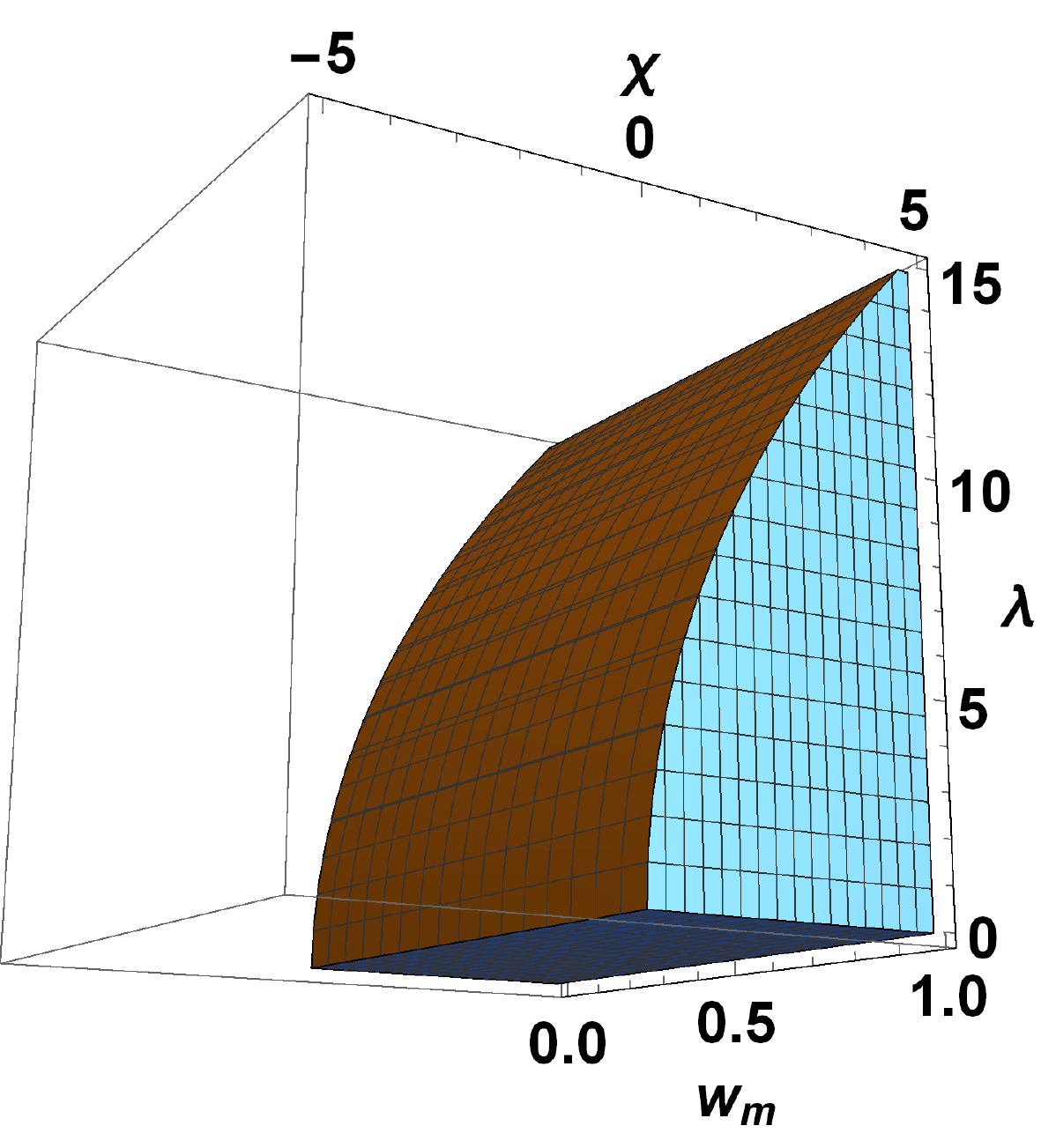} 
    \end{minipage}
     \caption{(a) The case where the critical point $O$ represents a stable cosmological scenario (left panel); (b) The situation where $O$ corresponds to a stable spiral evolution, due to complex eigenvalues (right panel).}
     \label{fig:unu}
\end{figure}

 
\noindent In this scenario, the effective equation of state for the dark energy field can be written as:
\begin{multline}
    w_{\rm eff}=\frac{1}{6 u^2 \left(x^2-1\right)^2+\sqrt{1-x^2}} \Big( -2 u \sqrt{3-3 x^2} x y w_{\rm m}+\sqrt{1-x^2} w_{\rm m}-y^2 w_{\rm m}+6 u^2 x^4-12 u^2 x^2+6 u^2+3 \lambda  u \sqrt{1-x^2} x^2 y^2
    \\-2 \lambda  u \sqrt{1-x^2} y^2+2 \sqrt{3} u \sqrt{1-x^2} x y-2 \sqrt{3} u \sqrt{1-x^2} x^3 y-2 \chi  \sqrt{1-x^2} x^2 y^2+x^2 y^2-y^2 \Big)\,.
\end{multline}
Next, the critical points are determined by considering that the RHS of Eqs.~(\ref{equu1})--(\ref{equu2}) are equal to zero, taking into account also the physical viability which requires: $0\leq \Omega_{m}=1-\Omega_{\phi}\leq 1$, $y\geq 0$, $1-x^2>0$ and the location of the corresponding critical points is the the real space. In the case of this specific power law coupling and for an exponential potential, the phase space structure has a reduced complexity, having only one critical point located at $O(x,y,u)=\left(0, 1,\frac{\lambda }{6}  \right)$, with the following eigenvalues
\begin{equation}
    \Xi_{O}=\left[-\frac{3 \left(\sqrt{\left(\lambda ^2+6\right) \left(\lambda ^2-48 \chi +6\right)}+\lambda ^2+6\right)}{2 \left(\lambda ^2+6\right)},\frac{3 \left(\sqrt{\left(\lambda ^2+6\right) \left(\lambda ^2-48 \chi +6\right)}-\lambda ^2-6\right)}{2 \left(\lambda ^2+6\right)},-3 \left(w_{\rm m}+1\right) \right]\,.
\end{equation}
This critical point corresponds to a de Sitter evolution, acting as a cosmological constant $ w_{\rm eff}= -1 $, implying the full domination of the tachyonic dark energy field over the matter component, $\Omega_m=1-\Omega_{\phi}=0$. For this critical point, we show in Fig.~\ref{fig:unu}, various regions for the model's parameters which are connected to different cosmological scenarios, corresponding to a stable and stable spiral evolution, respectively.

\subsection{Exponential coupling and inverse hyperbolic sine potential}
Here, we investigate the case where the coupling functional is represented by an exponential, $g(\phi)=g_0 e^{\alpha \phi}$, with $g_0$ and $\alpha$ constants. The potential energy associated with the present tachyonic dark energy model is considered to be an inverse hyperbolic sine, $V(\phi)= V_0 \sinh^{-\omega_1}(\omega_2 \phi)$, with $\omega_{1,2}$ constants. This type of potential is beyond the usual exponential type considered in many dynamical analysis and is motivated by the recent work in Ref.~\cite{Roy:2017uvr}, where the structure of the phase space for non--canonical fields with the potential energy beyond exponential type was investigated. The potential energy considered in this section, the inverse hyperbolic sine \cite{UrenaLopez:2000aj,Sahni:1999gb} represents a possible parameterization for the dark energy field which is associated to a second order polynomial in the dynamical equation for the dimensionless variable $\lambda$. As discussed in Ref.~\cite{Roy:2017uvr}, if we denote $f(\lambda)=\lambda^2(\Gamma-1)$ with $\Gamma$ defined in the relation in Eq.~(\ref{ecuatie_gamma}), then for this specific potential, the function $f(\lambda)$ obeys a second order polynomial parameterization $f(\lambda)=a \lambda^2+b \lambda + c$, with $a,b,c$ constant parameters. For the specific potential considered here, the inverse hyperbolic sine, we find that $f(\lambda)=\lambda^2/\omega_1-\omega_1 \omega_2^2$.

\begin{figure}[H]
    \begin{minipage}{0.35\textwidth}
        \centering
        \includegraphics[width=0.9\textwidth]{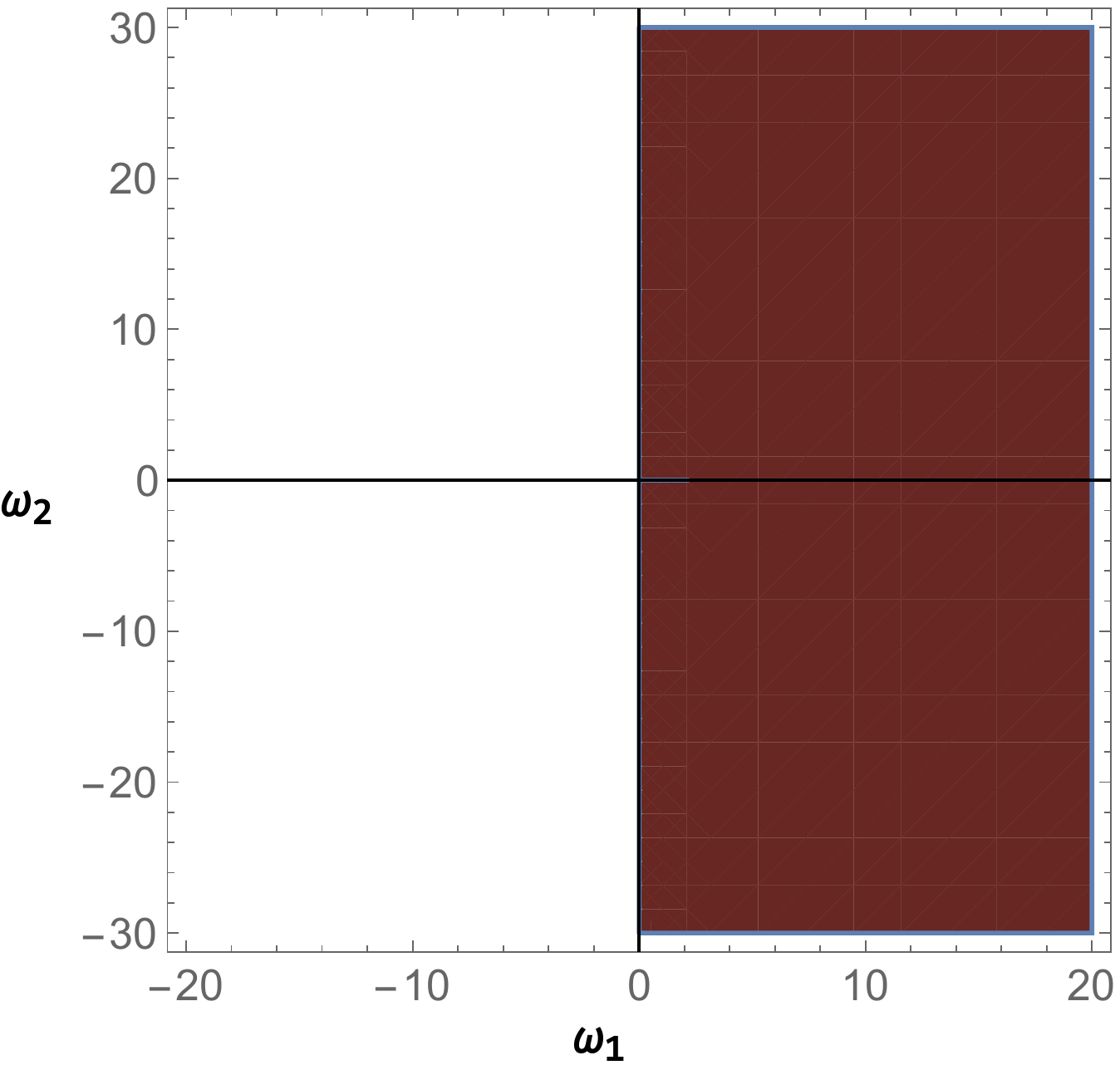} 
    \end{minipage}\hfill
    \begin{minipage}{0.35\textwidth}
        \centering
        \includegraphics[width=0.9\textwidth]{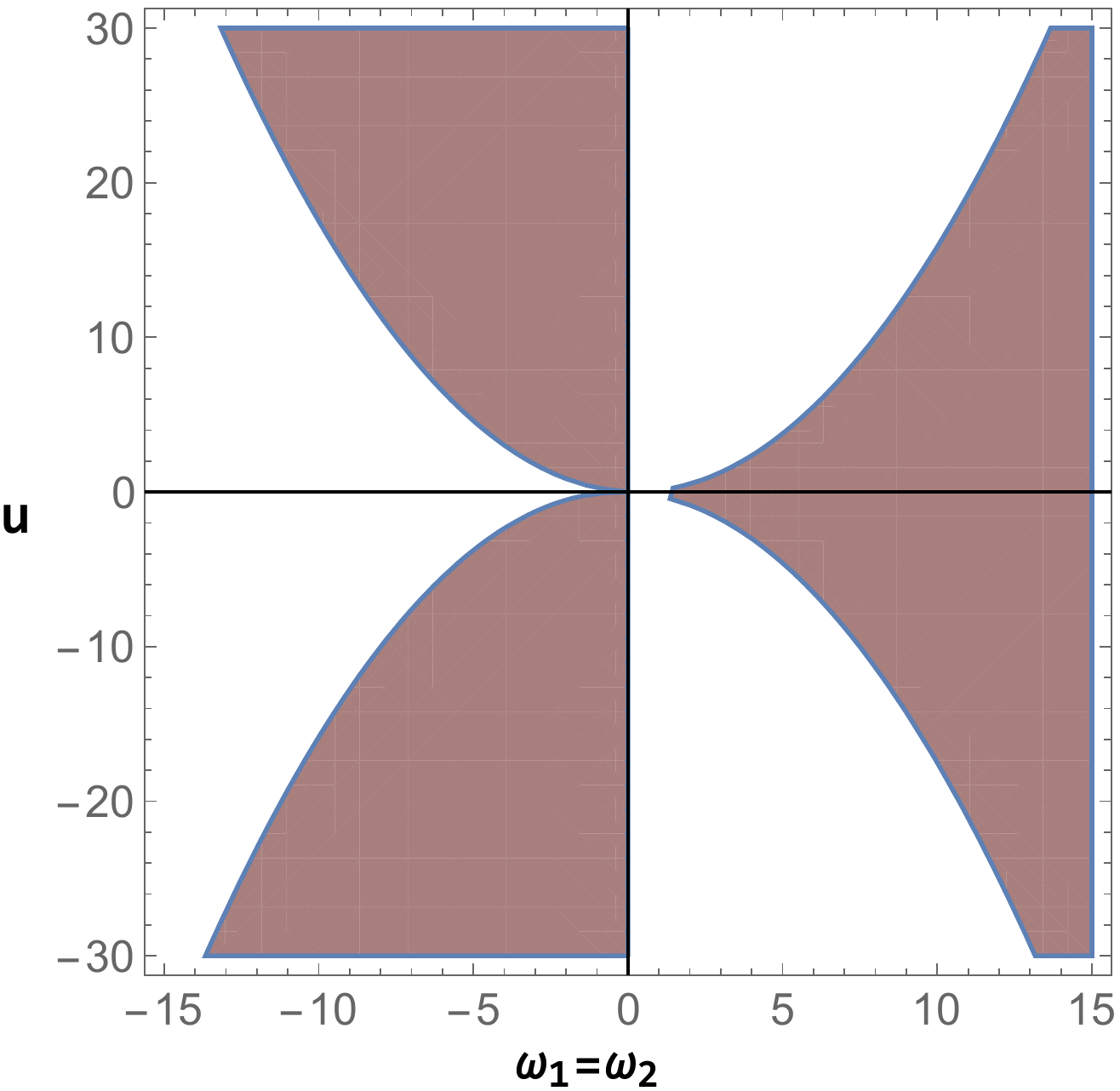} 
    \end{minipage}
     \caption{(a) The case where the critical line $A$ has a saddle cosmological behavior, considering $w_{\rm m}=0$, $u=0$, $\alpha=1$  (left panel); (b) The situation where $A$ corresponds to a saddle critical line, considering $w_{\rm m}=0$, $\omega_1=\omega_2$, $\alpha=1$ (right panel).}
     \label{fig:aaa}
\end{figure}

\begin{figure}[th]
    \begin{minipage}{0.35\textwidth}
        \centering
        \includegraphics[width=0.9\textwidth]{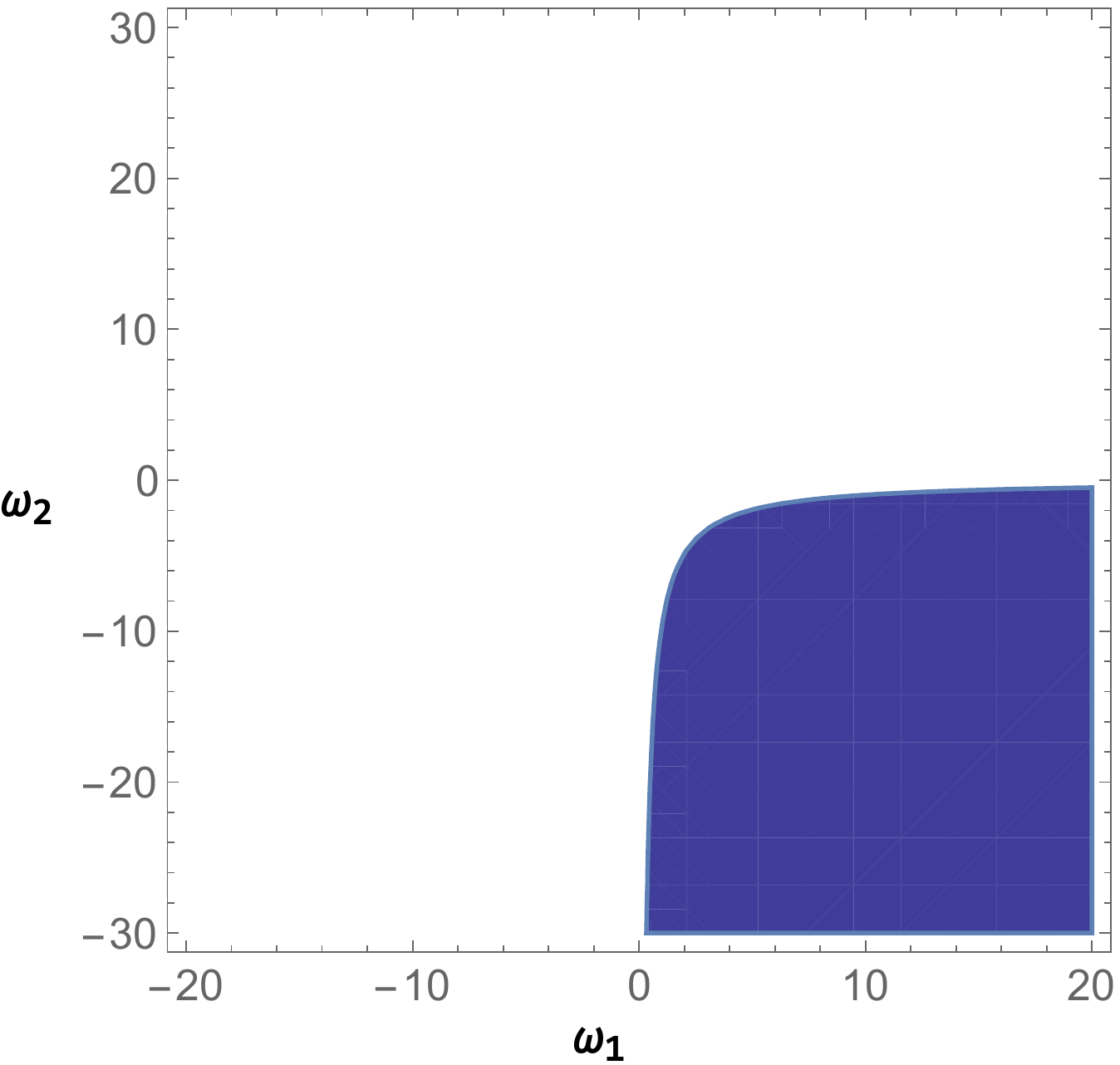} 
    \end{minipage}\hfill
    \begin{minipage}{0.35\textwidth}
        \centering
        \includegraphics[width=0.9\textwidth]{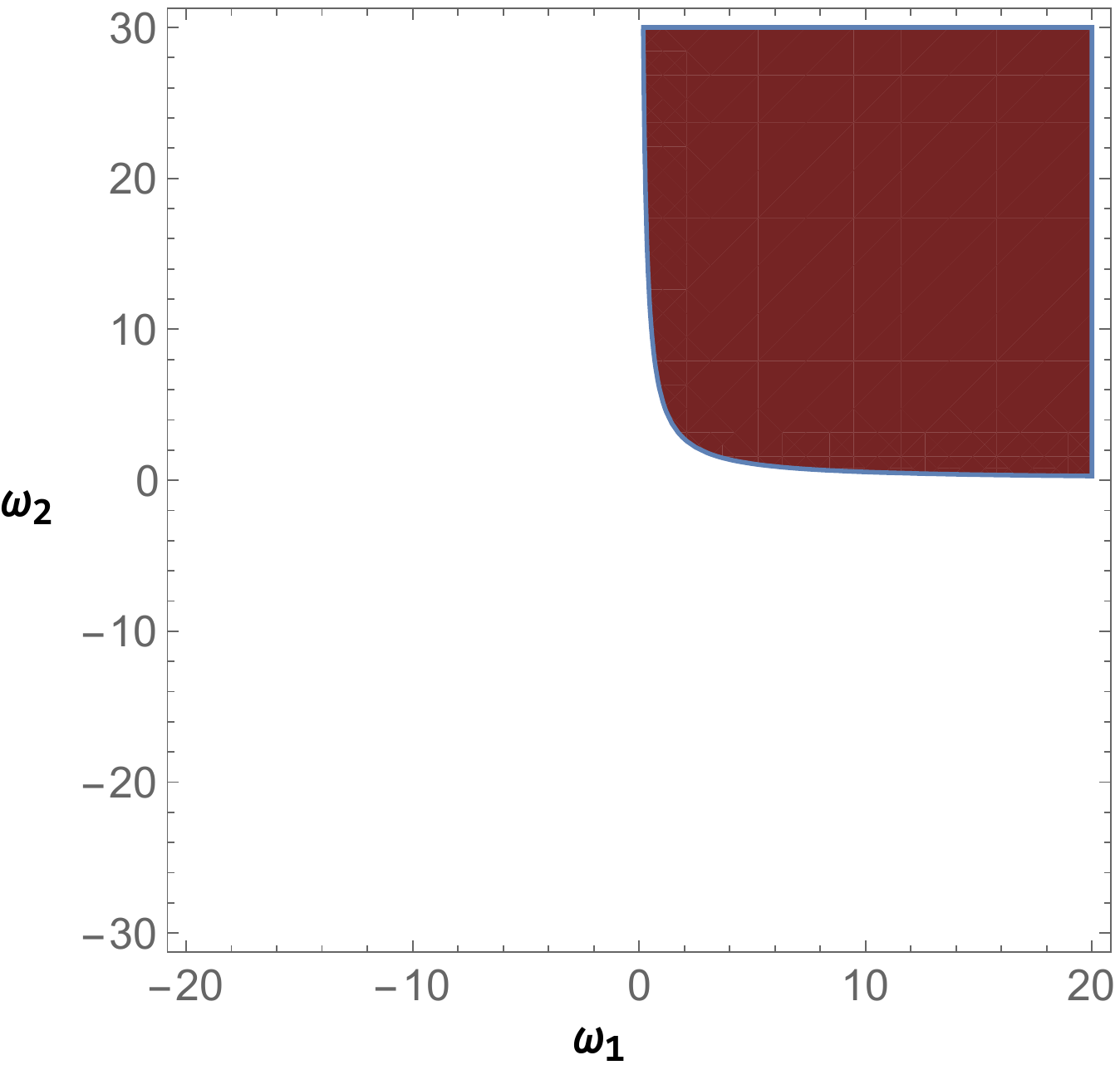} 
    \end{minipage}
     \caption{(a) The case where the critical point $B_{-}^{-}$ has a stable cosmological behavior, considering $w_{\rm m}=-0.001$, $\alpha=2$ (left panel); (b) The situation where $B_{+}^{+}$ corresponds to a stable scenario, considering $w_{\rm m}=-0.01$,  $\alpha=-1$ (right panel).}
     \label{fig:bbbaaa}
\end{figure}

In this case, the evolution of the dynamical system can be written as the following system of differential equations\begin{multline}
\frac{dx}{dN}=-\frac{1}{y \left(6 u^2 \left(x^2-1\right)^2+\sqrt{1-x^2}\right)}(x^2-1) \Big[-3 u w_{\rm m} \left(-6 u x^3 y+6 u x y+\sqrt{3-3 x^2} y^2+\sqrt{3} x^2-\sqrt{3}\right)
\\+3 \sqrt{3} u^2 x^2 \left(x^2-1\right) y^2 (2 \alpha -\lambda )+3 \sqrt{3} u \left(x^2-1\right) \left(\sqrt{1-x^2} y^2+1\right)+\sqrt{1-x^2} y \left(\sqrt{3} \lambda  y-3 x\right) \Big]\,,
\label{eq:12}
\end{multline}
\begin{multline}
\frac{dy}{dN}=-\frac{y}{2 \left(6 u^2 \left(x^2-1\right)^2+\sqrt{1-x^2}\right)} \Big[ 3 w_{\rm m} \left(2 u x \sqrt{3-3 x^2} y-\sqrt{1-x^2}+y^2\right)+6 u^2 \left(x^2-1\right)^2 \left(\sqrt{3} \lambda  x y-6\right)
\\+3 u \sqrt{1-x^2} y \left(2 \sqrt{3} x^3+x^2 y (2 \alpha -3 \lambda )-2 \sqrt{3} x+2 \lambda  y\right)-3 x^2 y^2+\lambda  x \sqrt{3-3 x^2} y-3 \sqrt{1-x^2}+3 y^2 \Big]\,,
\end{multline}
\begin{equation}
\frac{du}{dN}=\sqrt{3} \alpha  u x y\,,
\end{equation}
\newpage and finally the last equation for the dynamical system given by
\begin{equation}
\frac{d\lambda}{dN}=-\sqrt{3} x y \left(\frac{\lambda ^2}{\omega _1}-\omega _1 \omega _2^2\right)\,.\label{eq:16}
\end{equation}
The total effective equation of state is:
\begin{align}
    w_{\rm eff}&=\frac{1}{6 u^2 \left(x^2-1\right)^2+\sqrt{1-x^2}} \Big( -u \sqrt{1-x^2} y \left(2 \sqrt{3} x \left(w_{\rm m}-1\right)+2 \sqrt{3} x^3+x^2 y (2 \alpha -3 \lambda )+2 \lambda  y\right)
  \nonumber  \\
    &+w_{\rm m} \left(\sqrt{1-x^2}-y^2\right)+6 u^2 \left(x^2-1\right)^2+\left(x^2-1\right) y^2\Big)\,.
\end{align}

Next, as in the previous case, the critical points are obtained by considering that the RHS of the evolution equations in Eqs.~(\ref{eq:12})--(\ref{eq:16}) are equal to zero. For the potential energy of the inverse hyperbolic sine type, the critical points and the main physical properties are expressed in Table~\ref{table:que}. It should be noted that for dust matter $w_{\rm m}=0$, only the critical point $A$ exists in the phase space. As can be noted from this table, one can observe two main classes of critical points. The first class, denoted as $A$ represents a critical line which corresponds to a cosmological constant behavior, having an interrelation between the auxiliary variable $u$ associated to the coupling function $g(\phi)$ and the strength of the potential energy, embedded into the non--constant $\lambda$ variable. This epoch corresponds to a de Sitter universe, a critical line where the dark energy field dominates in terms of density parameters. Analyzing the corresponding eigenvalues, due to the presence of a zero eigenvalue, the linear stability method fails to provide a viable theoretical framework for determining the stability properties. Hence, for this critical line we can only argue on the specific cases where we have a saddle cosmological behavior, due to the presence of eigenvalues with both positive and negative real parts. Concerning this critical line, we display in Fig.~(\ref{fig:aaa}) various regions for the model's parameters which correspond to a saddle cosmological behavior, due to the existence of at least one eigenvalue with negative real part, and at least an eigenvalue with positive real part.

A second class of critical points displayed in Table~\ref{table:que} is represented by the $B_{i}^{j}$, $i,j=\left \{ +,- \right \}$ critical points which have a scaling cosmological behavior, an epoch in which the dark energy field acts as a matter component and mimics a matter dominated epoch. This type of solutions can in principle solve the cosmic coincidence problem. For these solutions, we have shown in Table~\ref{table:que} the locations in the phase space and the corresponding physical properties associated. From a physical point of view, the validity of these solutions implies that various  existence conditions are satisfied. These corresponds to the requirement that the critical points belong to the real space, and the corresponding density parameters are physically viable, $\Omega_{\phi}=1-\Omega_{m} \in \left[ 0,1 \right]$. Moreover, due to the definition of the dimensionless variables within this section, one should add the requirement that the $y$ variable is real and positive and $x \in \left( -1,1 \right)$ due to the form of the matter (dark energy) density parameter. From the expression of the $x$ variable presented in the table, it can be seen that the existence conditions imply that for the present tachyonic dark energy model $w_{\rm m} \in \left[-1, 0\right)$. Assuming that the matter component is embedded into the dark matter fluid, this implies that the pressure associated to the dark matter fluid is negative, an exotic situation which is not excluded by different cosmological observations~\cite{Thomas:2016iav, Yang:2016dhx, PhysRevD.88.127301,PhysRevD.71.047302}. Moreover, for the critical points $B_{\left[+,-\right]}^{-}$, the existence conditions imply that the product $\omega_{1}\omega_2$ is negative, while for the solutions $B_{\left[+,-\right]}^{+}$ we have an inverted situation, $\omega_{1}\omega_2>0$. Hence, the current tachyonic dark energy model might contribute to a solution of the cosmic coincidence problem due to the scaling solutions since it can recover matter and de Sitter cosmological epochs. The stability of the points $B_{\pm}^{\pm}$ depend on the sign of the eigenvalues. In general, the stability conditions for each point are very cumbersome since they depend on the parameters $\alpha, w_{\rm m}$ and $\omega_{1,2}$. Concerning the stability properties, we show in Fig.~\ref{fig:bbbaaa} various cases which corresponds to the stability associated to the $B_{-}^{-}$ and $B_{+}^{+}$ critical points, determining possible values of the $\omega_{1,2}$ parameters which result in stable scaling solutions. Point $A$ is a non-hyperbolic point and standard linear stability theory fails on describing any stability property of it. One can use other dynamical system techniques as centre manifold theory to study its stability (see~\cite{BAHAMONDE20181, Leon:2015via}). However, due to limited physical effects associated to the $A$ critical point, we relied our analysis only on linear stability methods, exploring the specific conditions where the stability corresponds to a saddle dynamical behavior. The present discussion can be adapted also for various potential types beyond exponential which have been studied in Ref.~\cite{Roy:2017uvr} with compatible results. 
\begin{table}[t!]
    \label{tab:table1}
    \begin{tabular}{|c|c|c|c|c|c|c|c|} 
 \hline
 Point & x & y & u & $\lambda$ & $\Omega_{\phi}$ & $w_{\rm eff}$ & Eigenvalues\\ 
 \hline\hline
 A& 0 & 1 & u & 6u & 1 & -1 & $0,-3 \left(w_{\rm m}+1\right),-\frac{3}{2} \pm \frac{\sqrt{3} \sqrt{\left(6 u^2+1\right) \omega _1 \left(\omega _1 \left(18 u^2-24 \alpha  u+3\right)-144 u^2+4 \omega _1^2 \omega _2^2\right)}}{2 \left(6 u^2+1\right) \omega _1}$ \\ 
 \hline
 $B_{+}^{-}$& $\sqrt{w_{\rm m}+1}$ & $-\frac{\sqrt{3} \sqrt{w_{\rm m}+1}}{\omega _1 \omega _2}$ & 0 & $-\omega _1 \omega _2$ & $\frac{3 \left(w_{\rm m}+1\right)}{\omega _1^2 \omega _2^2 \sqrt{-w_{\rm m}}}$ & $w_{\rm m}$ & $ -\frac{3 \alpha  \left(w_{\rm m}+1\right)}{\omega _1 \omega _2},-\frac{6 \left(w_{\rm m}+1\right)}{\omega _1}, \frac{3}{4} \left( w_{\rm m}-1\pm \Xi \right)$ \\ 
 \hline
 $B_{-}^{+}$& $-\sqrt{w_{\rm m}+1}$ & $\frac{\sqrt{3} \sqrt{w_{\rm m}+1}}{\omega _1 \omega _2}$ & 0 & $-\omega _1 \omega _2$ & $\frac{3 \left(w_{\rm m}+1\right)}{\omega _1^2 \omega _2^2 \sqrt{-w_{\rm m}}}$ & $w_{\rm m}$ & $-\frac{3 \alpha  \left(w_{\rm m}+1\right)}{\omega _1 \omega _2},-\frac{6 \left(w_{\rm m}+1\right)}{\omega _1}, \frac{3}{4}\left(w_{\rm m}-1 \pm \Xi \right) $ \\ 
 \hline
 $B_{-}^{-}$& $-\sqrt{w_{\rm m}+1}$ & $-\frac{\sqrt{3} \sqrt{w_{\rm m}+1}}{\omega _1 \omega _2}$ & 0 & $\omega _1 \omega _2$ & $\frac{3 \left(w_{\rm m}+1\right)}{\omega _1^2 \omega _2^2 \sqrt{-w_{\rm m}}}$ & $w_{\rm m}$ & $ \frac{3 \alpha  \left(w_{\rm m}+1\right)}{\omega _1 \omega _2},-\frac{6 \left(w_{\rm m}+1\right)}{\omega _1}, \frac{3}{4}\left(w_{\rm m}-1 \pm \Xi \right)$ \\
 \hline
  $B_{+}^{+}$& $\sqrt{w_{\rm m}+1}$ & $\frac{\sqrt{3} \sqrt{w_{\rm m}+1}}{\omega _1 \omega _2}$ & 0 & $\omega _1 \omega _2$ & $\frac{3 \left(w_{\rm m}+1\right)}{\omega _1^2 \omega _2^2 \sqrt{-w_{\rm m}}}$ & $w_{\rm m}$ & $\frac{3 \alpha  \left(w_{\rm m}+1\right)}{\omega _1 \omega _2},-\frac{6 \left(w_{\rm m}+1\right)}{\omega _1}, \frac{3}{4}\left(w_{\rm m}-1 \pm \Xi \right) $ \\ 
 \hline
\end{tabular}
  \caption{The critical points in the case of exponential coupling and inverse hyperbolic sine potential. The auxiliary variable $\Xi$ used in the description of the eigenvalues for the various critical points is equal to: $\Xi=\frac{\sqrt{\omega _1^4 \omega _2^4 \left(-w_{\rm m}\right) \left(w_{\rm m}+1\right) \left(17 \omega _1^2 \omega _2^2 w_{\rm m}^2+14 \omega _1^2 \omega _2^2 w_{\rm m}-96 \left(-w_{\rm m}\right){}^{3/2}+48 \left(-w_{\rm m}\right){}^{5/2}+48 \sqrt{-w_{\rm m}}+\omega _1^2 \omega _2^2\right)}}{\omega _1^3 \omega _2^3 \sqrt{-w_{\rm m} \left(w_{\rm m}+1\right)}}$}
\label{table:que}
\end{table}

\section{Conclusions}\label{sec:4}
In this paper, we have considered a new dark energy model in the teleparallel equivalent of general relativity, based on modifications due to tachyonic fields which are non--minimally coupled with the torsion scalar and its boundary term. In this approach, the boundary term is related to the divergence of the torsion vector and the torsion scalar reproduces the same theory as GR at the level of the field equations. After finding the corresponding field equations for this tachyonic dark energy model, we analyzed the effects of the non--minimal coupling by employing the linear stability theory. In the first scenario, we considered the case where the coupling function is represented by a power law dependence, and the potential energy term corresponds to an exponential. In this case, we observed that the structure of the phase space has a reduced complexity. The critical points which are present corresponds to a cosmological constant behavior. Hence, in this case one notices that the evolution of the dynamical system can explain the current accelerated expansion of the present Universe due to the cosmological constant behavior of the system at the critical points. However, due to the reduced complexity of the phase space and without the presence of the scaling solutions, the cosmic coincidence problem cannot be alleviated.  
\par 
A second cosmological scenario was also considered by taking into account that the non--minimal coupling functional has an exponential type parameterization. Concerning the potential energy term in the action, the study considered that the potential is beyond the usual exponential type found in many dynamical constructions in scalar tensor theories. As can be noted from the previous section, the dynamical equation associated to the potential term involves a second order polynomial parameterization. In this specific case in the analysis, the potential corresponds to an inverse hyperbolic sine. For the second cosmological scenario the phase space have a 4-dimensional structure and a richer complexity. As can be noted from the previous section, we have shown that the second cosmological scenario analyzed can reproduce the known evolution of the Universe and solve the cosmic coincidence problem due to the existence of scaling solutions. In these critical points, the dark energy field mimics a matter era due to the specific form the corresponding effective equation of state. However, the existence conditions associated to the scaling solutions imply the existence of an exotic warm dark matter fluid having a limited negative pressure, a comportment not ruled out by present astrophysical observations \cite{Thomas:2016iav,Aghanim:2018eyx}. Furthermore, the rest of the critical points in the phase space corresponds to a cosmological constant behavior and can explain the current accelerated expansion of the Universe with a constant equation of state.
\par 
One can then say that the current dark energy model constructed in the teleparallel equivalent of general relativity modified by a tachyonic field non--minimally coupled with a boundary term represents a potentially realistic model in scalar tensor theory which might solve the cosmic coincidence problem and the nature of the dark energy phenomenon, a feasible tachyonic prototype. This offers one possible alternative avenue to tachyonic fields to compliment the curvature-based work in the literature \cite{Quiros:2009mz,Aguirregabiria:2004xd,Bagla:2002yn,Fang:2010zze,Otalora:2013tba,Otalora:2014aoa}.

\begin{acknowledgments}
The JLS and SB would like to acknowledge networking support by the COST Action GWverse CA16104. This article is based upon work from CANTATA COST (European Cooperation in Science and Technology) action CA15117, EU Framework Programme Horizon 2020. SB is supported by Mobilitas Pluss N$^\circ$ MOBJD423 by the Estonian government. M. Marciu acknowledge partial support by the project 29/2016 ELI-RO from the Institute of Atomic Physics, Bucharest--Magurele.
\end{acknowledgments}

\bibliographystyle{Style}
\bibliography{bibtele}

\end{document}